\documentclass{ws-procs9x6-cpt16}
\begin{document}

\newcommand{\refeq}[1]{(\ref{#1})}
\def\etal {{\it et al.}}

\title{Testing Gravity on Accelerators}

\author{T.\ Kalaydzhyan}

\address{Department of Physics, University of Illinois at Chicago, Chicago, IL 60607, USA}

\address{Department of Physics and Astronomy, Stony Brook University\\
Stony Brook, NY 11794, USA}

\begin{abstract}
The weak equivalence principle (WEP) is one of the cornerstones of the modern theories of gravity, stating that the trajectory of a freely falling test body is independent of its internal structure and composition. Even though WEP is known to be valid for the normal matter with a high precision, it has never been experimentally confirmed for relativistic matter and antimatter. We make an attempt to constrain possible deviations from WEP utilizing the modern accelerator technologies. We analyze the (absence of) vacuum Cherenkov radiation, photon decay, anomalous synchrotron losses, and the Compton spectra to put limits on the isotropic Lorentz violation and further convert them to the constraints on the difference between the gravitational and inertial masses of the relativistic electrons/positrons. Our main result is the 0.1\% limit on the mentioned difference.\\
\end{abstract}

\bodymatter

\phantom{}\vskip10pt\noindent
The gravitational field of the Earth (or Sun or other distant massive celestial objects) around an accelerator can be considered homogeneous and described by an isotropic metric for a static weak field, $ds^2 = \mathcal{H}^2 dt^2 - \mathcal{H}^{-2}(dx^2 + dy^2 + dz^2)$, where $\mathcal{H}^2 = 1+2\Phi$, and $\Phi$ is the gravitational potential, defining the acceleration of free-falling bodies.
For a massive particle (in our case electron/positron) with gravitational mass $m_{e,g}$, one can write the gravitational potential as\cite{Kalaydzhyan:2015ija} $\Phi_m = \Phi \,{m_{e,g}}/{m_e}\,, \mathcal{H}_m^2 \equiv 1+ 2\Phi_m$, which will modify the dispersion relation of the positron with momentum $\textbf{p}$ and energy $\mathcal{E} \gg m_e$ and the relation between energy and mass,\cite{Kalaydzhyan:2015ija}
\begin{equation}
\textbf{p}^2 = (1+2\kappa) \left(\mathcal{E}^2 - m_e^2 \right),\quad\qquad
{\cal E} = \frac{m_e \mathcal{H}^{-1}\mathcal{H}_m}{\sqrt{1-\mathcal{H}^{4}\mathcal{H}_m^{-4}\textbf{v}^2}}, \label{energy}
\end{equation}
where $\kappa = 2 |\Phi| \Delta m_e / m_e$, $\Delta m_e = m_{e,g} - m_e$, and $\mathbf{v}$ is the velocity of the particle ($c = 1$). We consider no change in the photon dispersion relation due to strong existing constraints on the variation of the speed of light. 
We obtain several experimental limits on $\kappa$ that will be translated to $\Delta m_e / m_e$ when taking into account variation of the (solar) potential $\Phi = \Phi_\odot$.

The on-shell emission of a photon by an electron or positron in the vacuum, so-called vacuum Cherenkov radiation, is normally forbidden kinematically. However, in the presence of the nontrivial modification of the dispersion relation (\ref{energy}) with $\kappa < 0$, the energy-momentum conservation condition allows such a process. In other words, the electron (positron) is allowed to move faster than light at a certain energy. The energy threshold ${\cal E}_\mathrm{th}$ 
is given then by\cite{Altschul:2007tn} ${\cal E}_\mathrm{th} = {m_e}/\sqrt{- 2 \kappa}$.
Due to the high emission rate, a particle above ${\cal E}_\mathrm{th}$ will be rapidly slowed down to the threshold energy through the photon radiation. For instance, the electrons/positrons at LEP at CERN, with the energies ${\cal E}=104.5$ GeV and the arbitrarily chosen threshold energy ${\cal E}_\mathrm{th}=100$ GeV would be decelerated to the subluminal speeds just within 1.2~cm of travel\cite{Hohensee:2009zk} (compare to, e.g., $\sim 6$~km distance between LEP accelerating RF systems). Since this was never observed, ${\cal E}_\mathrm{th}> 100$ GeV and $\kappa > \kappa^- = -1.31\times 10^{-11}$.
 
As another standard textbook example, decay of a photon into an electron-positron pair is also forbidden kinematically. However, at $\kappa > 0$ it becomes possible. The threshold on the photon energy $\omega_\mathrm{th}$ is given by\cite{Hohensee:2009zk} $\omega_\mathrm{th} = \sqrt{{2}/{\kappa}} \,m_e$,
where we assumed for simplicity that the electron's dispersion relation is modified in the same way as the positron's, since there are no precise limits on the gravitational mass of the ultrarelativistic electron either. Following Ref.\ \refcite{Hohensee:2009zk}, we consider isolated photon production with an associated jet, $p\bar p \rightarrow \gamma + \mathrm{jet} + X$, as measured by the D0 detector at Fermilab Tevatron collider at the center-of-mass energy $\sqrt{s}=1.96$~TeV. 
The possible photon decay process is very efficient and leads to a fast energy loss. As an example, 300~GeV photons with an energy 1\% above threshold would decay after traveling an average distance of only 0.1~mm (for comparison, the photons should travel a minimal distance of 78~cm in order to be measured by the central calorimeter of the D0 detector). As shown in Ref.\ \refcite{Hohensee:2009zk}, the hypothetical photon decay at 300~GeV would lead to a deficit in the photon flux much larger than that allowed by the difference between QCD and experimental data.
This leads to the right bound $\kappa < \kappa^+ = 5.80\times 10^{-12}$. 

Taking the two-sided bound $\kappa^- < \kappa < \kappa^+$ for two potentials, $\Phi$ and $\Phi + \Delta\Phi$ (e.g., both the vacuum Cherenkov radiation and photon decay were absent during the experiment), one can easily derive 
\begin{equation}
\kappa^- - \kappa^+ < - 2 \Phi_\odot \frac{\Delta d_{SE}}{d_{SE}} \frac{\Delta m_{e}}{m_e} < \kappa^+ - \kappa^-\,, \label{newlimits}
\end{equation}
where $\Delta d_{SE}$ is the variation of the distance between Sun and Earth, $d_{SE}$, due to the eccentricity of the Earth's orbit, and obtain\cite{Kalaydzhyan:2015ija}
$\left| \Delta m_{e}/{m_e}\right| < 0.0389$, i.e., a 4\% limit on the possible deviation.

In an ultrarelativistic case, $|\textbf{v}| \approx 1$, the ratio ${\cal E}/m_e$ from Eq.~(\ref{energy}) modifies the synchrotron radiation power $P = 2 e^2 {\dot {\mathbf v}}^2 (\mathcal E / m_e)^4 /3$ of a circular accelerator by an amount $\Delta P$, such that $\Delta P / P  = -4 \kappa \gamma^2$,
where $\gamma \equiv 1/\sqrt{1-\textbf{v}^2}$. One can relate the fractional deviation in the masses and the fractional uncertainty in the measured synchrotron radiation power in two experiments,
\begin{equation}
\left|\frac{\Delta m_{e}}{m_e}\right| < \frac{|\Delta P / P|_{1} + |\Delta P / P|_{2}}{8 \gamma^2 |\Phi_\odot \Delta d_{SE}[\mathrm{AU}]|}\,.\label{limit2}
\end{equation}
Following Ref.\ \refcite{Altschul:2009xh}, we consider the beam energy measurements of the LEP Energy Working Group for the LEP 2 programme in the last few years of LEP operation. Analysis of their data on the synchrotron losses collected at different moments of time leads to the limit $\left| \Delta m_{e}/{m_e}\right| < 0.0013$, i.e., 0.1\%, for both electrons and positrons.\cite{Kalaydzhyan:2015qwa}

As a possible future improvement of the limits, one can study the Compton edge for the scattering of laser photons of energy $\omega_0$ and incident angle $\theta_0$ off high energy electrons/positrons. If the Compton edge for the photon is measured in two experiments at its nominal position $\omega_{max}$ within uncertainties $\Delta \omega_1$ and $\Delta \omega_2$, then\cite{Kalaydzhyan:2015ija}
\begin{align}
\left|\frac{\Delta m_e}{m_e}\right| < \frac{\Delta \omega_1 + \Delta \omega_2}{\omega_{max}}\cdot\frac{m_e^2 (1+x)^2}{4{\cal E}^2 |\Delta \Phi|}\,,
\end{align}
where $x \equiv 4{\cal E} \omega_0\sin^2{(\theta_0/2)}/m_e^2$ is a kinematic parameter defined by the experimental setup. Continuous laser Compton scattering experiments at the future ILC and CLIC accelerators with estimated sensitivity $|\kappa| \sim 10^{-13}$ can improve our best limits by an order of magnitude.\cite{Kalaydzhyan:2015ija}

\section*{Acknowledgments}
This work was supported in part by the U.S.\ Department of Energy 
under contracts No.\ DE-FG-88ER40388 and DE-FG0201ER41195.

\end{document}